\documentstyle[12pt]{article}
\begin{document}
\hfill SINP/TNP/93-18\break\smallskip
\hfill hep-th/9310146\bigskip
\begin{center}
{\large{\bf A Generalized Gauge Invariant Regularization\\
of the Schwinger Model}}\\\medskip
G. Bhattacharya, A. Ghosh and P. Mitra\footnote{
e-mail mitra@saha.ernet.in}\\
Saha Institute of Nuclear Physics\\
Block AF, Bidhannagar\\
Calcutta 700 064, INDIA
\vfill
Abstract
\end{center}
The  Schwinger  model is studied with a new one - parameter class
of gauge invariant regularizations that generalizes the usual
point - splitting or Fujikawa schemes. The spectrum is found to be
qualitatively unchanged, {\it except for a limiting value of  the
regularizing   parameter,  where  free  fermions  appear  in  the
spectrum.}
\vfill\eject
\section{Introduction}
The  Schwinger  model  \cite{Sch},  namely the theory of massless
quarks interacting with an abelian gauge field in two-dimensional
spacetime, has been extensively studied over the  years  and  has
provided  theorists  a  lot of insight into the phenomena of mass
generation and confinement \cite{LS,CKS}.  The  quark  disappears
from the physical spectrum in this model,  leaving  only  a  free
massive particle associated with the gauge field. Exact solutions
are available for various operators and Green functions.

The  regularization  underlying  the  conventional  study  of the
Schwin-ger model is such that the physical mass  of  the  particle
becomes equal to ${1\over\sqrt\pi}$ times the bare gauge coupling
constant.  This  regularization maintains the gauge invariance of
the theory although a mass is  generated  for  the  gauge  field.
Other regularizations that give up gauge invariance have recently
been studied \cite{PM} and lead to different physical results
-- the  quark  gets  liberated  in  that situation much as in the
closely  related  chiral Schwinger model \cite{JR}. However, even
if gauge invariance is not abandoned, it is possible to make  the
regularization  more  flexible, for example in the context of the
Fujikawa regularization scheme. The
nature of the solution is not qualitatively changed --  only  the
relation  between  the  physical  mass  and  the bare coupling is
generalized. In one sense, the theory is not changed at all, for
there is only one physical quantity in the picture -- the mass  of
the  particle -- and it is a dimensional object in two dimensions,
so  that  its {\it value} is not relevant. In another sense, this
regularization gives rise to a  new  relation  between  fermionic
bilinears  and  bosons  so  that  one effectively has a different
bosonization  scheme. This flexibility allows an unusual limit to
be taken, whereby the physical mass can be made zero. This  opens
up  a  new  scenario in this model. It is to the consideration of
the new regularized  version  of  the  Schwinger  model  and  the
special limiting case that the present paper is devoted.

The  plan  of  the  paper is as follows. We first explain how the
regularization of the Schwinger model allows an extra flexibility
in the effective action of the model. This implies a  generalized
expression for the anomaly in the axial current of the theory. It
is  shown  that  the  usual  form  of the fermion operator of the
Schwinger model allows the current to be constructed  in  such  a
way that this generalized expression is obtained for the anomaly.
The  gauge field equation of motion is then satisfied only if the
physical mass is related to the  gauge  coupling  constant  in  a
specific  way  depending  on  the  regularization. This fixes the
effective action of the theory in terms  of  the  gauge  coupling
constant  and  the  regularization. The quark-antiquark potential
following from this  effective  action  is  worked  out  and  the
special limiting case investigated.
\section{ Operator solution of equations of motion}
The  Schwinger  model  is  described  by  the  Lagrangian density
\cite{Sch}
\begin{equation}
{\cal   L}=\overline{\psi}(i\partial\!\!\!/+eA\!\!\!/)\psi
-{1\over  4}F^{\mu\nu}F_{\mu\nu},
\end{equation}
where the indices  take  the  values  0,1  corresponding   to   a
(1+1)$-$
dimensional spacetime and the notation is standard.
In two dimensions we can always set
\begin{equation}
A_\mu  =  -{\sqrt{\pi}\over   e}(\widetilde\partial_\mu\sigma   +
\partial_\mu\widetilde\eta),\label{2}
\end{equation}
where, \begin{equation}
\widetilde\partial_\mu = \epsilon_{\mu\nu}\partial^\nu\end{equation}
with $\epsilon_{01}=+1$ and
$\sigma,\widetilde\eta$  are scalar fields.

In  this  section we shall restrict ourselves to the Lorentz gauge,
where from (\ref{2}) we see that the field  $\widetilde\eta$  can
be  taken  as  a  massless  field  with $\Box\widetilde\eta=0$. We can then
introduce its dual through
\begin{equation}
\widetilde\partial_\mu\eta(x)=\partial_\mu\widetilde\eta(x).\end{equation}
These  massless  fields  have  to  be  regularized because in two
dimensions the two point function  of  a  massless  scalar  field
diverges  \cite{LS}.  We  shall not need the explicit form of the
regularization here.

The Dirac equation in the presence of the gauge field is
\begin{equation}
\left[i\partial \!\!\!/ + eA\!\!\!/\right]\psi(x)=0.
\end{equation}
It is easy to check that this equation is satisfied by
\begin{equation}
\psi    (x)=:~e^{i\sqrt{\pi}\gamma_5   \left[\sigma   (x)   +\eta
(x)\right]}:\psi^{(0)}(x),\label{7}
\end{equation}
where, $\psi^{(0)}(x)$ is a free fermion field
satisfying $i{\partial\!\!\!/}\psi^{(0)}(x)=0$.

We can calculate  the  gauge  invariant current using the point -
splitting regularization. While constructing  a  gauge  invariant
bilinear  of fermions which in the limit of zero separation would
give the  usual  fermion  current,  we  generalize  slightly  the
conventional construction \cite{Sch}. We take
\begin{equation}
J_\mu^{reg}(x)=\lim_{\epsilon\rightarrow 0}\left[\overline\psi(x+\epsilon)
\gamma_\mu~:e^{ie\int^{x+\epsilon}_x dy^\mu\big\{ A_\mu(y)-a
\partial^\nu F_{\mu\nu}(y)\big\} }:\psi(x) - {\rm v.e.v.}\right]
\end{equation}
where $a$  is  an  arbitrary parameter. The term in  the  exponent
containing   this   parameter   is   new   and   represents   our
generalization of the usual regularizing phase factor. This term
preserves gauge invariance, Lorentz invariance and
even the linearity of the theory,
but its natural occurrence has not been realized before.
Now using (\ref{2}) and (\ref{7}) together with
\begin{equation}
F_{\mu\nu}={\sqrt\pi\over e}\epsilon_{\mu\nu}\Box\sigma\label{F}
\end{equation}
we obtain the current  which,  upto an overall
wavefunction renormalization,  is equal to
\begin{eqnarray}
J_\mu^{reg}(x) &\approx & ~:\overline{\psi^{(0)}}(x)\gamma_\mu\psi^{(0)}(x):-
i\sqrt\pi\lim_{\epsilon\rightarrow 0}\langle 0\mid
\overline{\psi^{(0)}}(x+\epsilon)\gamma_\mu
\bigl[(\gamma_5\epsilon \cdot\partial
\nonumber\\& +&\epsilon\cdot\widetilde\partial)(\sigma+\eta)
+a\epsilon\cdot\widetilde\partial\Box\sigma\big]\psi^{(0)}(x)\mid 0\rangle\\
&=& ~:\overline{\psi^{(0)}}(x)\gamma_\mu\psi^{(0)}(x):\nonumber\\&-&
{1\over \sqrt \pi}\big[{\epsilon_\mu
\epsilon_\nu-\widetilde\epsilon_\mu\widetilde\epsilon_\nu\over
{\epsilon^2}}\widetilde\partial^\nu(\sigma+\eta)+ a{
\epsilon_\mu\epsilon_\nu\over {\epsilon^2}}\widetilde\partial^\nu
\Box\sigma\big],
\end{eqnarray}
where we have used the identity
\begin{equation}
\langle 0\mid\overline{\psi^{(0)}}_\alpha  (x+\epsilon)\psi_\beta (x)\mid
0\rangle =-i{\epsilon\!\!\!/_{\beta\alpha}\over 2\pi\epsilon^2}.
\end{equation}
Now we take  the  symmetric limit {\it i.e.\/} average over the
point splitting directions $\epsilon$ and finally obtain
\begin{equation}
J_\mu^{reg}(x)=-{1\over \sqrt\pi}\widetilde\partial_\mu(\phi+\sigma+a\Box
\sigma+\eta),
\end{equation}
where $\phi$ is a free massless bosonic field satisfying
\begin{equation}
-{1\over{\sqrt\pi}}\widetilde\partial_\mu\phi=:\overline\psi^
{(0)}(x)\gamma_\mu\psi^{(0)}(x):\end{equation}   and  thus  representing  the
bosonic equivalent of the free   fermionic   field   $\psi^{(0)}$
\cite{Cole}. This field too has to be understood to be regularized.
We find
\begin{eqnarray}
J_{\mu 5}^{reg}(x)&=&\epsilon_{\mu\nu}J^\nu_{reg}(x)\\
&=&-{1\over{\sqrt\pi}}
\partial_\mu(\phi +\eta +
\sigma +a\Box\sigma),
\end{eqnarray}
so that the anomaly is
\begin{equation}
\partial^\mu  J_{\mu 5}^{reg}=
-{1\over{\sqrt\pi}}
\Box(\phi +\eta +
\sigma +a\Box\sigma).\label{j}\end{equation}

Note now that Maxwell's equation with sources, {\it viz.},
\begin{equation}
\partial_\nu F^{\nu\mu} + eJ^\mu_{reg} =0,\end{equation}
can be converted to the pair of equations
\begin{equation}
\left[~\Big(1+{ae^2\over \pi}\Big)\Box~+~{e^2\over \pi}~\right]\sigma=0
\label{m}
\end{equation}
and \begin{equation}
\phi +\eta=0.\end{equation}

The second equation relating two massless free fields
will be satisfied in a weak sense by imposing a subsidiary condition
\begin{equation}
(\phi + \eta)^{(+)}\mid phys\rangle =0\label{s}\end{equation}
to select out a physical subspace of states.
We shall also ensure that $\phi + \eta$
creates only states with zero norm by taking $\eta$ to
be  a negative metric field, {\it i.e.}, by taking its
commutators to have the ``wrong'' sign.  The  subsidiary  condition
then also serves
to separate out a subspace with nonnegative metric as usual.

We  see  from  (\ref{m})  that $\sigma$ is a massive free field, as
expected.
The  only  difference  from the usual case is the presence of the
factor $(1+{ae^2\over\pi})$.
This  implies  that  the
spectrum  of the theory as regularized here is the same as in the
usual  case  with  the  mass  scaled  down by a factor $\sqrt{(1+
{ae^2\over\pi})}$.
\section{
Effective action of QED$_2$}
In the previous section we regularized the current directly as an
operator product of fermion fields. The same regularized current
will  now   be  obtained  from an effective action which we shall
construct through a Fujikawa regularization.

The effective action is defined
by the following functional of the abelian gauge field
$A_\mu$:
\begin{equation}
e^{i\Gamma[A]}=\int{\cal  D}\psi{\cal  D}\overline\psi   e^{i\int d^2x
\overline\psi iD\!\!\!/\psi},\label{1}
\end{equation}
where, $D_\mu = \partial_\mu-ieA_\mu$.
Notice that by virtue of (\ref{2}) and the identity
\begin{equation}
\gamma^\mu\epsilon_{\mu\nu} = \gamma^5\gamma_\nu,\end{equation}
which holds in two dimensions, we can write
\begin{equation}
D\!\!\!/=\partial\!\!\!/+i\sqrt{\pi}\partial\!\!\!/\widetilde\eta +
i\sqrt\pi\gamma_5\partial\!\!\!/\sigma.\end{equation}
It is easy to see  that  the  transformations
\begin{equation}
\psi'=e^{i\sqrt\pi(\widetilde\eta - \gamma_5\sigma)}\psi,
\end{equation}
\begin{equation}
\overline{\psi}'=\overline{\psi}e^{-i\sqrt{\pi}(\widetilde\eta +\gamma_5
\sigma)},\end{equation}
decouple  the  gauge field from the fermions and the classical
action becomes free, {\it i.e.},
\begin{equation}
\overline{\psi}i{D\!\!\!/}\psi =\overline{\psi}'i{\partial\!\!\!/}\psi'.
\end{equation}

But in the quantum theory this decoupling from the  action  leads
to  a  non-trivial change in the fermionic measure, which is
related to the chiral anomaly.
To calculate the   Jacobian  we   must proceed    through
infinitesimal transformations of the fermionic fields in the path
integral.
So we define
\begin{eqnarray}
\psi'_\delta & = &(1+i\sqrt\pi\delta\widetilde\eta-i\sqrt\pi
\gamma_5\delta\sigma)\psi,\nonumber\\
\overline{\psi}'_\delta &=&
\overline{\psi}(1-i\sqrt\pi\delta\widetilde\eta-i\sqrt\pi
\gamma_5\delta\sigma),\label{3}
\end{eqnarray}
leading to,
\begin{equation}
\overline{\psi}D\!\!\!/\psi=\overline{\psi}'_\delta
[\partial\!\!\!/+i\sqrt\pi\partial\!\!\!/
(\widetilde\eta-\delta\widetilde\eta) + i\sqrt\pi\gamma_5\partial\!\!\!/
(\sigma-\delta\sigma)]\psi'_\delta.
\end{equation}
The Jacobian corresponding to this transformation, defined by
\begin{equation}
{\cal D}\psi{\cal D}\overline{\psi}=J^\delta{\cal D}\psi'_\delta {\cal D}
\overline{\psi}'_\delta ,\end{equation} {\it viz.},
\begin{equation}
J^\delta_{reg}=e^{2i\sqrt{\pi}\int d^2x~
tr~\gamma_5\delta\sigma(x)},\end{equation}
is regularized to
\begin{equation}
J^\delta_{reg}=e^{2i\sqrt{\pi} ~Tr~\gamma_5\delta\sigma(x)
e^{t({D\!\!\!/}^r)^2}},\label{zeta}\end{equation}
where
$D^r_\mu$  is  a regularizing antihermitian
differential operator, $Tr$ stands for the
full trace and the limit $t\to 0^+$ is to be taken.
Fujikawa chose the operator $D^r_\mu$  to be
the Euclidean Dirac operator \cite{Fuji}.
Other choices, {\it e.g.}, in \cite{Har}, correspond to different
regularizations. To calculate the trace, it is convenient to take
a plane wave basis. Then the exponent in (\ref{zeta}) simply gets
multiplied  by  a  factor  ${a_1\over 4\pi}$, which is defined as
follows. First the
Dirac operator  is  continued  to  the  Euclidean  space; after
evaluating the trace it is finally continued back  to
the  Minkowski  space.  Hence  in  the  following  calculation
we  have to use Euclidean gamma   matrices
(although  the  same notation is used as for
Minkowski gamma matrices). $a_1$ is given by
\begin{equation}
a_1=(D\!\!\!/^r)^2- (D^r)^2
\end{equation}

The regularization considered in the previous section corresponds to
choosing the regularizing Dirac operator to be
\begin{equation}
{D^r}_\mu=\partial_\mu       -ieA_\mu
-iae\partial_\nu   F_{\nu\mu}.\end{equation}
 By  (\ref{2}),  \begin{equation}{D\!\!\!/}^r  =\partial\!\!\!/  +i\sqrt\pi
\partial\!\!\!/\widetilde\eta    -\sqrt\pi   \gamma_5\partial\!\!\!/(\sigma
+a\Box\sigma),\end{equation} which gives
\begin{equation}
a_1 =\sqrt\pi\gamma_5\Box(\sigma +a\Box\sigma).\label{4}
\end{equation}

The calculation of the effective action goes as follows.
By the transformations (\ref{3}) we can write (\ref{1}) as
\begin{equation}
e^{i\Gamma[\sigma,\widetilde\eta]}=J^\delta_{reg}~e^{i\Gamma[\sigma
-\delta\sigma,\widetilde\eta -\delta\widetilde\eta]}.
\end{equation}
Thus,
\begin{equation}
\delta\Gamma[\sigma]={1\over i}\log~J^\delta_{reg}[\sigma]=
2\sqrt\pi\int d^2x\; tr\,\gamma_5
\delta\sigma(x){a_1\over  4\pi}\end{equation}  and  \begin{equation}
{\delta\Gamma\over
{\delta\widetilde\eta}}=0.\end{equation}
Using (\ref{4})  and finally integrating to a finite $\sigma(x)$, we
get
\begin{equation}
\Gamma[\sigma]={1\over 2}\int d^2x~\left[\sigma\Box\sigma +a\Box\sigma\Box
\sigma\right].
\end{equation}
Finally, using the inverse of (\ref{2}) and
(\ref{F}), we obtain the effective action
\begin{equation}
\Gamma[A]=\int  d^2x~\left[{e^2\over 2\pi}\widetilde\partial\cdot A{1\over
\Box}\widetilde\partial\cdot A - {ae^2\over
4\pi}F_{\mu\nu}F^{\mu\nu}\right].
\end{equation}
This effective action can be used to calculate the fermionic
currents $eJ_\mu ={\delta\over {\delta A^\mu}}~\Gamma[A]$ and
\begin{eqnarray}
J_5^\mu &=&\epsilon^{\mu\nu}J_\nu\nonumber\\
&=& {e\over\pi}\epsilon^{\mu\nu}\left[A_\nu +a\partial^\rho F_{\nu\rho}
-{1\over\Box}\partial_\nu\partial\cdot A\right],\end{eqnarray}
from which we find the
anomaly equation to be
\begin{eqnarray}
\partial_\mu                                J_5^\mu &=&{e\over
2\pi}\epsilon_{\mu\nu}[F^{\mu\nu} + a\Box F^{\mu\nu}]\\
&=& -{1\over     {\sqrt\pi}}\Box(\sigma
+a\Box\sigma),
\end{eqnarray} which is
consistent with (\ref{j}) when the subsidiary  condition (\ref{s}) is
imposed.
\section{The bosonization of QED$_2$}
If  we  make  the above  effective  action  local  by  introducing  an
auxiliary field $\Sigma$ and insert the kinetic energy  term  for
the gauge field, we obtain
the bosonized action of $QED_2$  generalized as above:
\begin{equation}
S_B=\int d^2x~\left[-{1\over 4}\Big(1+{ae^2\over  \pi}\Big)F^2
+{e^2\over                    2\pi}A^2                   +{1\over
2}\partial_\mu\Sigma\partial^\mu\Sigma    -    {e\over
 \sqrt{\pi}}A^\mu\partial_\mu\Sigma\right].
\end{equation}

The effective action leads to a Hamiltonian through standard
constraint analysis as follows. First, the canonical momenta have
to be defined. The momenta corresponding to $A_0, A_1$ and $\Sigma$
are respectively
\begin{equation}
\Pi_0=0,\label{p}
\end{equation}
\begin{equation}
\Pi^1=\Bigl(1+{ae^2\over\pi}\Bigr)(\partial_0A_1-\partial_1A_0),
\end{equation}
\begin{equation}
\Pi_\Sigma=\dot\Sigma-{e\over\sqrt\pi}A_0.
\end{equation}
(\ref{p}) is recognized to be a constraint.
Using all these equations, we obtain the Hamiltonian
\begin{equation}
{\cal   H}={(\Pi^1)^2\over   2(1+{ae^2\over\pi})}    +    {1\over
2}\Pi_\Sigma^2 + {e\over\sqrt\pi}A_0\Pi_\Sigma + \partial_1A_0\Pi^1 +
{e^2\over 2\pi}A_1^2 + {1\over 2}\Sigma'^2 -  {e\over\sqrt\pi}\Sigma'
A_1.
\end{equation}
The  consistency  of  (\ref{p})  under
time  evolution  by  this  Hamiltonian   requires   a   secondary
constraint
\begin{equation}
G\equiv\partial_1\Pi^1 - {e\over\sqrt\pi}\Pi_\Sigma=0.\label{g}
\end{equation}
There  are no further constraints, and it can be checked that the
Poisson brackets of  (\ref{p}) and (\ref{g}) with one another
vanish,  so  that  the constraints are {\it first class}. This is
natural, as we have taken care to maintain  gauge  invariance  in
the  effective  action. As usual, then, we have to fix a gauge to
remove gauge  degrees  of  freedom.  It  is  convenient  here  to
consider the physical gauge conditions
\begin{equation}
\Sigma=A_0=0.
\end{equation}
(In  the next section we shall use a different kind of gauge fixing.) In
the present gauge, the Hamiltonian simplifies to
\begin{equation}
{\cal   H}={(\Pi^1)^2\over   2(1+{ae^2\over\pi})}   +   {e^2\over
2\pi}A_1^2 + {\pi\over 2e^2}(\Pi^{1\prime})^2,
\end{equation}
which may be converted to the familiar form
\begin{equation}
{\cal H}= {1\over 2}\Pi_\Phi^2 + {1\over 2}\Phi'^2  +  {1\over  2}
{e^2\over \pi+ae^2}\Phi^2
\end{equation}
by the redefinitions
\begin{equation}
\Phi={\sqrt\pi\over e}\Pi^1,~~~\Pi_\Phi=-{e\over\sqrt\pi}A_1.
\end{equation}
This  shows that the physical spectrum of the model contains just
a massive boson with mass ${e\over\sqrt{\pi+ae^2}}$.
\section{Confinement and deconfinement of  quarks}
Let us investigate the nature of the force mediated by the  gauge
field of this theory between two quarks. First,
in   the   presence   of   two  static external quarks
($q\overline q$-pair) of
charge $Q$ at $\pm {L\over 2}$, the charge density is modified to
\begin{eqnarray}
J_q^0  (t,x^1)&=&{Q\over  e}\Bigl[\delta(x^1-{L\over  2})   -\delta(x^1
+{L\over  2})\Bigr]  +{e\over  \pi}A_0  -{e\over  \pi}{\partial_0\over
\Box}\partial.A +{ae\over \pi}\partial_1 F_{01}\nonumber\\
&=&J^0 -{1\over {\sqrt\pi}}\partial_1\chi,\end{eqnarray}
where,
\begin{equation}
\chi={\sqrt\pi\over e}Q\theta(x^1 +{L\over  2})\theta(x^1
-{L\over  2}).\end{equation}  So  the  Lagrangian  density in the presence of
these external quarks can be written as
\begin{equation}
{\cal L_Q}=-{1\over 4}\Bigl(1+{ae^2\over \pi}\Bigr)F^2 +{e^2\over {2\pi}}A^2
+{1\over 2}\partial_\mu\Sigma\partial^\mu\Sigma -{e\over {\sqrt\pi}
}\partial_\mu\Sigma                 A^\mu                -{e\over
{\sqrt\pi}}\widetilde{\partial}\cdot A\chi.
\end{equation}
{}From a   constraint analysis similar to the one in section 4,  we   get   the
corresponding Hamiltonian density in the physical gauge to be
\begin{equation}
{\cal   H_Q}={1\over   2}\widetilde\Pi^2_\Phi +{1\over   2}\widetilde\Phi'^2
+{1\over 2}{e^2\over {\pi+ae^2}}{(\widetilde\Phi -\chi)}^2,
\end{equation}
where $\widetilde\Pi_\Phi=\Pi_\Phi$ and $\widetilde\Phi=\Phi +2\chi$.

The difference in ground state energies between ${\cal H_Q}$
and ${\cal H}$ can be calculated to be
\begin{equation}
E_Q -E ={1\over 2}\int dx^1\left[{e^2\over \pi+ae^2}\chi^2 +({e^2\over
\pi      +ae^2})^2\chi(\partial_1^2      -{e^2\over      \pi
+ae^2})^{-1}\chi\right].
\end{equation}
Hence the potential between the quark-antiquark pair is
\begin{equation}
V(L)={1\over  2}\cdot{Q^2\over  {e^2/\pi}}\cdot{e\over  {\sqrt{\pi
+ae^2}}}\big[1 -e^{-{eL\over {\sqrt{\pi +ae^2}}}}\big],
\end{equation}
which  is constant for large $L$, indicating the screening of the
charges as in the usual version of the Schwinger model.  However,
in the limit of massless gauge fields
$ae^2\rightarrow\infty,~~
V(L)=0$,  i.e.  the  (external) quarks become free. This is to be
contrasted with the limit $e\rightarrow 0$ of the  usual  version
of the Schwinger model or simply the free electromagnetic theory,
where $V(L)={1\over 2}Q^2L$, so that there is a  linearly  rising
confining  potential. Thus we are  led to expect deconfinement in
the limit $ae^2\rightarrow\infty$. The existence of free massless
fermions in this  limit  of  the  theory  can  be  understood  by
noticing  that  the  ordinarily  massive  boson  present  in  the
spectrum becomes massless in this limit and  {\it  this  massless
free boson can be regarded as the bosonized version of a massless
free  fermion  field.}  For further evidence of deconfinement, we
consider the behaviour of the dynamical quarks.

The bosonized  action  can  be  used  to calculate the two point
correlation function of the fermions \cite{prop}.  First,  from  the
equation
\begin{equation}
(i\partial\!\!\!/ +eA\!\!\!/)G_F(A;x,y)=\delta^2 (x,y),
\end{equation}
we  can  express  $G_F(A;x,y)$  in terms of the free fermion
Green function $S_F(x,y)$ by perturbative expansion in $e$:
\begin{equation}
G_F(A;x,y)=e^{-ie\int d^2z A_\mu(z)j^\mu(z;x,y)}~S_F(x,y),
\end{equation}
where                              \begin{equation}
j_\mu(z;x,y)=(\partial_\mu^z
+\gamma_5\widetilde\partial_\mu^z)\left[D_F(z-x) -D_F(z-y)
\right]\end{equation}
and formally\begin{equation}
D_F(x)=-\int {d^2p\over (2\pi)^2}{e^{-ip\cdot x}\over  p^2+i0}.
\end{equation}
$D_F$ should include a regularization to take  care  of  the
infrared divergence. But in this context
where $D_F$ appears only in a differentiated form, there is no divergence.

The  two point function  $G_F(x,y)$  can  be  calculated  from
$G_F(A;x,y)$
by integrating out the background
field $A$. Thus,
\begin{equation}
G_F(x,y)=e^{-{ie^2\over       2}\int       d^2z_1\int       d^2z_2
j^\mu(z_1;x,y)G_{\mu\nu}(z_1,z_2)j^\nu(z_2;x,y)}~S_F(x,y),\label{gf}
\end{equation}
where
\begin{equation}
G_{\mu\nu}(x,y)={m^2\over e^2/\pi}\Bigl[g_{\mu\nu}
+\bigl\{\alpha{e^2\over        \pi}({1\over        m^2}       +{1\over
\Box})-1\bigr\}{\partial_\mu\partial_\nu\over
\Box}\Bigr]\Delta_F(x,y;m^2)\end{equation}
is   the   two   point function of the gauge field defined with a
gauge fixing term $-{1\over 2\alpha}\int ~d^2x~(\partial\cdot A)^2$
added  to  the action. Furthermore,
\begin{equation}
\Delta_F(x;m^2)=-\int {d^2p\over
(2\pi)^2}{1\over  p^2-m^2+i0} e^{-ip\cdot
x},\end{equation}
the scalar field propagator.

(\ref{gf}) simplifies to
\begin{equation}
G_F(x,y)=\exp\biggl[i\int {d^2p\over {(2\pi)^2}}\Bigl[-{\pi  m^2\over  {p^2(p^2
-m^2)}} +{\alpha e^2\over p^4}\Bigr](1 -e^{-ip.(x-y)})\biggr]S_F(x,y).
\end{equation}
The form of this two point function
makes  it  difficult to  say  anything
definite  about  the large separation behaviour of the Green
function. It is therefore desirable
to  calculate  the  gauge  invariant  two  point  function
\cite{giprop}.

The  gauge invariant two point function is the vacuum expectation
value of the gauge invariant bilocal operator
\begin{equation}
T(x,y) = \psi(x) \overline\psi(y)
:e^{ie\int^{x+\epsilon}_x dy^\mu\big\{ A_\mu(y)-a
\partial^\nu F_{\mu\nu}(y)\big\} }:.
\end{equation}
The  term  -  $a\partial_\nu  F^{\mu\nu}$  has  been  included to
maintain the identity
\begin{equation}
J^\mu_{reg}(x)=-\lim_{\epsilon\rightarrow 0}~tr~\gamma^\mu
[T(x,x+\epsilon) - {\rm v.e.v.}].\end{equation}

It is clear that if we can express the
line  integral  as  a  volume  integral,
the  gauge  invariant  two  point  function  will   be  given  by
(\ref{gf}) with
only  a  modification in the current density $j_\mu$.
This is achieved by the use of the identity
\begin{equation}
\int^y_x d\xi_\mu V^\mu (\xi)=\int d^2z V^\mu(z)s_\mu(z;x,y)
\end{equation}
where $s_\mu(z;x,y)=(y-x)_\mu \int^1_0 dt\delta^{(2)}(\xi(t; x,y)-z)$,
$V_\mu$ is an arbitrary vector and
the path of integration $\xi_\mu(t;x,y)=(y-x)_\mu   t
+   x_\mu$ is  taken  as a straight line.
Hence the gauge invariant two  point  function  in  a  background
gauge field is given by
\begin{equation}
G_F{}^{g.i.}(A;x,y)=\exp\biggl[-ie\int d^2z
A_\mu(z)[j^\mu(z;x,y)-\widetilde s^\mu(z;x,y)]\biggr]S_F(x,y),
\end{equation}
where $\widetilde s^\mu= s^\mu +a(\Box s^\mu-\partial^\mu\partial\cdot s).$
Now $\partial_\mu(j^\mu-\widetilde s^\mu)=0$, so the
gauge dependent part of $G_{\mu\nu}$ does not contribute in
$G_F{}^{g.i.}$. Hence on integrating out the gauge field we get
\begin{equation}
G_F{}^{g.i.}(x,y)=\exp\biggl[-{ie^2\over 2}\int\!\!\int
(j_\mu-\widetilde s_\mu)\big\{ {
m^2\over e^2/\pi}g^{\mu\nu}\Delta_F \big\}
(j_\nu-\widetilde s_\nu)\biggr]~S_F(x,y).
\end{equation}
As $(\partial_\mu +\gamma_5\widetilde\partial_\mu)
(\partial^\mu +\gamma_5\widetilde\partial^\mu)=0$, the  diagonal  term  in
$j_\mu$ does not contribute in the phase factor.
Furthermore, using the identity $\partial_\mu^z s_\nu (z;x,$ $y)=
\partial_\nu^z s_\mu (z;x,y)$
we  can  see  that $\widetilde s_\mu$ can be replaced by $s_\mu$.
The detailed calculation yields
\begin{eqnarray}
G_F{}^{g.i.}(x,y)&=&\exp\biggl[-i\pi\int {d^2p\over (2\pi)^2}[-{2\over p^2}
+{(x-y)^2\over                             [p.(x-y)]^2}]{m^2\over
p^2-m^2}(1\nonumber\\ &-& e^{-ip.(x-y)})\biggr]~
S_F(x,y),
\end{eqnarray}
which shows that in the limit  $ae^2\rightarrow\infty$,
$G_F{}^{g.i.}\rightarrow  S_F$.  This  is  a  clear indication of the
deconfinement  of  quarks  in   this  limit.

Unfortunately  the pole structure of the propagators is not clear
for finite values of $a$ and  we  shall  have  to  consider other
arguments  to understand why the spectrum depends so crucially
on whether $a$ is finite or
infinite. The confinement of quarks in the usual Schwinger  model
is  understood  by imposing the subsidiary condition. Since the
operator solutions (\ref{7}) and (\ref{2}) for $\psi$ and $A$    do  not
commute  with  the  operator  $\phi  +  \eta$,  they  create both
physical and unphysical  states  from  the  vacuum.  It  is  more
convenient to make a gauge transformation and pass to the new set
of solutions
\begin{equation}
\psi' (x)=:~e^{i\sqrt{\pi}\big(\gamma_5 [\sigma (x) +\eta (x)]
+ \widetilde\eta (x)\bigr)}:\psi^{(0)}(x),\end{equation}
\begin{equation}
A'_\mu=-{\sqrt\pi\over e}\widetilde\partial_\mu\sigma.\end{equation}
Now according to \cite{Man},
\begin{equation}
\psi^{(0)} (x)\propto ~:e^{i\sqrt{\pi}\bigl(\gamma_5 \phi (x)
+\widetilde\phi(x)\bigr)}:,\label{Man}\end{equation}
so that by virtue  of  the  subsidiary  condition (\ref{s}),  $\psi'  (x)$  is
essentially $:e^{i\sqrt{\pi}\gamma_5 \sigma (x)}:$,  apart  from
cluster   -violating  operators  which  reduce  to  c-numbers  in
irreducible sectors \cite{LS}. These expressions clarify why there is no
fermion  in the spectrum for finite $a$.

For  infinite  $a$,  on the other hand, $\sigma(x)$ is a massless
field. One can then introduce its dual $\widetilde\sigma(x)$ through
\begin{equation}
\partial_\mu\widetilde\sigma(x)=\widetilde\partial_\mu\sigma(x),\end{equation}
and perform a gauge  transformation  with  it  to  construct  new
operator solutions of the equations of motion
\begin{equation}
\psi'' (x)=:~e^{i\sqrt{\pi}\bigl[\gamma_5 [\sigma (x) +\eta (x)]
+\widetilde\sigma(x)+ \widetilde\eta(x)\bigr]}:\psi^{(0)}(x),\end{equation}
\begin{equation}
A''_\mu=0.\end{equation}
Clearly, after the unphysical fields present  in  the  expression
for  $\psi''$  are  replaced  by  c-numbers,  what  is  left is a
representation of a free massless fermion in terms  of
$\sigma$ and its dual, {\it i.e.}, the analogue of (\ref{Man}) with
$\phi$ replaced by $\sigma$. This is how a fermion appears in the limit
of infinite $a$.
\section{Conclusion}
In  this  paper  we  have  looked  at  the Schwinger model with a
somewhat generalized regularization. First we point -  split  the
current which is formally defined as the product of two fermionic
operators.   Schwinger   has   prescribed  the  insertion  of  an
exponential of a line integral of the gauge  field  to  make  the
product gauge invariant. However, his choice was only one of many
possible choices. We have inserted an extra factor which involves
the  field  strength  of  the  gauge field and therefore does not
interfere with the gauge invariance of the product.  It  is  here
that  our  parameter  $a$ enters. Obviously, this is not the most
general gauge invariant regularization possible in this approach,
but the introduction of more complicated factors makes the theory
difficult to solve. With our  regularization,  the  equations  of
motion  of  the  Schwinger  model  can be converted to free field
equations exactly as in the usual case, with only the mass of the
scalar field altered by a factor involving the new parameter. The
conventional indefinite metric treatment  has  been  used  and  a
subsidiary condition imposed to separate out a physical space.

There  is one question which may arise in the reader's mind. Have
we, in changing the regularization, changed the model? To be more
specific,  the introduction of $A_\mu -a\partial^{\nu}F_{\mu\nu}$
instead  of  just  $A_\mu$  in  the  phase  factor  entering  the
point-split current may be suspected to amount to the addition of
an  extra  interaction  of  the  form  $-aj^{\mu}  \partial^{\nu}
F_{\mu\nu}$. This is not really the case,  as  the  equations  of
motion of the Schwinger model itself are satisfied. The change is
only   in  the  definition  of  fermion  bilinears  as  composite
operators and this is well known to have a  lot  of  flexibility.
Formally,  in  the  limit  $\epsilon\to 0$, the phase factor does
reduce to unity, so that the definition of the bilinears  adopted
in this paper is by no means unnatural.

After  the  operator  treatment,  a  Fujikawa  regularization  is
constructed in such a way that it gives the same  result  as  the
generalized  point  splitting procedure. This is used to find the
effective action of the theory. The nonlocal terms  present  here
can  be  recast in a local form as usual by the introduction of a
new scalar field, {\it viz.}, the  bosonized  equivalent  of  the
fermion field. A Hamiltonian analysis is carried out to establish
the  physical content of the theory, which may not be immediately
clear from the operator solution in an indefinite metric space.

The  question  of  confinement  has been discussed in detail. The
potential between external quarks has been  calculated.  A  gauge
invariant  propagator  has  been studied. Last but not least, the
operator solution itself has been scrutinized  with  the  aim  of
finding  the  fermion  content  of  the theory. All these studies
point in one direction: there are no fermions in the spectrum for
finite values of the parameter $a$, as is to be expected from the
usual treatment of the Schwinger model  corresponding  to  $a=0$,
but  when  this  parameter goes to infinity, a fermion reappears,
{\it i.e.\/}, deconfinement occurs.  This  is  not  difficult  to
understand  at  all. When the parameter $a$ goes to infinity, the
scalar which is  ordinarily  massive  becomes  massless.  In  two
dimensions  massless scalars are equivalent to massless fermions,
which explains the appearance of the fermion in this limit.

This  paper  is  limited  to the Schwinger model, but it is clear
that the ambiguity in point  splitting  regularization  that  has
been  exploited  here  exists in other models as well. While four
dimensional models may be difficult to handle, we  hope  to  deal
with  other  abelian and nonabelian models in two dimensions in a
separate publication.

We hope that investigations of this kind will throw more light on
the  not  too well understood phenomenon of quark confinement and
its  connection  with  details  of  regularization.  Hand  waving
arguments about confinement and deconfinement are almost all that
there is in four dimensions. The dependence of these phenomena on
regularization  schemes  clearly  indicates  the  need  for  more
quantitative investigations. Much work has of course been done on
the  lattice,  but  that is only one regularization. It has to be
generalized.

\end{document}